# Wafer-scale selective area growth of GaN hexagonal prismatic nanostructures on c-sapphire substrate

X. J. Chen,[1,*] J. S. Hwang,[2] G. Perillat-Merceroz,[3,4] S. Landis,[3] B. Martin,[3] D. Le Si Dang,[2] J. Eymery,[1] and C. Durand[1]

[1] Equipe mixte CEA-CNRS-UJF "Nanophysique et semiconducteurs", CEA, INAC, SP2M, NPSC, 17 rue des Martyrs, 38054 Grenoble cedex 9, France

[2] Equipe mixte CEA-CNRS-UJF "Nanophysique et semiconducteurs", Institut Néel-CNRS, 25 rue des Martyrs, BP 166 Fr-38042 Grenoble Cedex 9, France

[3] CEA, LETI, Minatec Campus, 38054 Grenoble Cedex 9, France

[4] CEA, INAC, SP2M, LEMMA, 17 rue des Martyrs, 38054 Grenoble cedex 9, France

Email: xiaojun.chen@cea.fr

* Author to whom any correspondence should be addressed.

ABSTRACT

Selective area growth of GaN nanostructures has been performed on full 2" c-sapphire substrates using $Si_3N_4$ mask patterned by nanoimprint lithography (array of 400 nm diameter circular holes). A new process has been developed to improve the homogeneity of the nucleation selectivity of c-oriented hexagonal prismatic nanostructures at high temperature (1040 °C). It consists of an initial GaN nucleation step at 950 °C followed by ammonia annealing before high temperature growth. Structural analyses show that GaN nanostructures are grown in epitaxy with c-sapphire with lateral overgrowths on the mask. Strain and dislocations are observed at the interface due to the large GaN/sapphire lattice mismatch in contrast with the high quality of the relaxed crystals in the lateral overgrowth area. A cathodoluminescence study as a function of the GaN nanostructure size confirms these observations: the lateral overgrowth of GaN nanostructures has a low defect density and exhibits a stronger near band edge (NBE) emission than the crystal in direct epitaxy with sapphire. The shift of the NBE positions versus nanostructure size can be mainly attributed to a combination of compressive strain and silicon doping coming from surface mask diffusion.

1. Introduction

The group-III nitride semiconductor nanostructures, such as quantum wells (QWs), quantum dots (QDs) and more recently nanowires (NWs) have attracted considerable interests, especially for photonic applications as light emitting diodes and lasers [1, 2, 3, 4]. Significant efforts have been made to control the growth of GaN compound nanostructures in terms of size and density to get homogeneous structural and optical properties. Selective area growth (SAG) using patterned masks have been especially demonstrated to be an effective way to localize the growth of nitride nanostructures. This method is essentially used in metal organic vapor phase epitaxy (MOVPE) with patterned dielectric mask ($SiO_2$ or $Si_3N_4$) to control the position and the size of GaN nanostructures as stripes [5], pyramids [5] and rods [6], and more recently in molecular beam epitaxy (MBE) using Ti patterned masks [7]. In MOVPE, SAG experiments are mostly reported on GaN layer template grown on c-sapphire substrate and usually result in pyramid-shaped nanostructures [8-12]. However, prismatic-shaped GaN structures have been observed for stripes and hexagonal nanostructures in SAG on c-sapphire [13, 14]. Indeed, nitridated sapphire surface favours the growth of N-polar GaN crystal [15]. Such polarity is particularly interesting to promote the growth of hexagonal GaN wires [16, 17]. This motivates the SAG of GaN nanostructures directly on c-sapphire that gives the opportunity to grow well organized c-oriented hexagonal GaN wires [18]. Nevertheless, the control of homogeneity of the nucleation selectivity of SAG GaN nanostructures on c-sapphire substrate remains an issue to be addressed. In addition, some more fundamental issues like the structural and optical properties of SAG nanostructures grown on sapphire have to be carefully studied especially due to the large lattice mismatch between sapphire and GaN.

In this work, we present a new method based on a temperature dependent two-step growth process to achieve the homogeneity of the nucleation selectivity growth of c-oriented GaN hexagonal prismatic-shaped nanostructures directly on patterned c-sapphire substrates. The patterning is obtained by nanoimprint lithography (NIL), which exhibits relevant advantages to other nanoscale patterning methods (for example electron beam lithography) with no charge effect coming from insulating materials and full wafer surface patterning at low cost and time-consumption. The first part of the paper will detail how the GaN nanostructure shape and the SAG nucleation homogeneity are related to the growth temperature, then it will be shown that hexagonal prismatic-shaped GaN SAG nanostructures can be managed with a high homogeneity by a two-step temperature growth process combined with an annealing procedure. Finally, thanks to the use of sapphire substrate without any intermediate GaN template layer, the optical properties of single GaN hexagonal nanostructures will be studied with respect to the structural properties to understand the influence of the nanostructure size.

## 2. Experimental methods

The patterning of 2" c-sapphire substrates was carried out by NIL using a Si mold with a uniform square array of circular pillars. Fig. 1 shows the three main steps to prepare the patterned substrates. Firstly (see Fig. 1(a)), a 5 nm-thick $Si_3N_4$ dielectric mask was deposited by plasma enhanced chemical vapour deposition at 300 °C. Hexamethyldisilazane (HMDS) was then spin coated to promote adhesion of the imprint resist. Secondly, a 225 nm-thick resist (NEB22 from Sumitomo) was spread and soft-baked at 110 °C for 2 minutes. The imprint process was done at 110 °C using a contact-force of 10 kN in a EVG®520HE nanoimprint equipment [19]. The residual resist layer at the bottom of the holes was removed by a short oxygen plasma treatment to reach the $Si_3N_4$ surface (see Fig. 1(b)). Thirdly, the $Si_3N_4$ mask was etched by using SF6 plasma followed by resist stripping to obtain a uniform patterning on the whole 2" wafers with a regular array of circular holes of 400 nm in diameter spaced by 1.1 µm (see the Fig. 1(c) and the AFM image in Fig. 1(d)). The choice of the circular shape of the openings is motivated by simplicity arguments in order to avoid taking into account the anisotropy of the pattern shape and its influence on the growth in terms of facet evolution.

The patterned substrates were then transferred into a MOVPE 3 x 2" close-coupled showerhead reactor (Aixtron CCS®) to carry out the growth at high pressure (800 mbar) using a flow of 8000 sccm of $N_2$ as carrier gas. Before growth, an in-situ pre-treatment of the patterned sapphires was done at high temperature (~ 1200 °C) using two steps: first a bake under H2 atmosphere during 20 min to clean the surface followed by a surface nitridation using ammonia flow (200 sccm) for 400 s. It has been shown that this preparation leads to the formation of a spontaneous thin AlN layer [16] driving the growth of N-polar GaN crystals [15, 17]. For GaN deposition, trimethylgallium (TMGa) and ammonia (NH$_3$) were used as precursors for the III and V materials. The V/III molar ratio was set to 16 corresponding to an injection flux of 135 μmol/min and 2.230 mmol/min for TMGa and NH$_3$ and the growth temperature has been set between 900 and 1040 °C. The choice of these growth conditions allows avoiding the lateral growth as explained in Ref.16.

The morphology of the as-grown samples was characterized by field emission scanning electron microscopy (FE-SEM). The structural properties were analyzed by X-ray diffraction (XRD) using the Co Kα (λ ~ 0.17902 nm) radiation and by 400 keV transmission electron microscopy (TEM) equipped with selected area electron diffraction (SAED). The optical properties of single GaN nanostructures were measured by cathodoluminescence spectroscopy (CL) at 5 K using 30 keV electron beam high voltage with about 1 nA current.

## 3. Results and Discussion

*3.1. Growth study: effect of temperature*

*3.1.1. Single-step temperature growth*

Selective growths on patterned sapphires were performed at 900, 950, 1000 and 1040 °C and Fig. 2(a)-(d) shows the corresponding 45°-tilted SEM images. At 900 °C (see Fig. 2(a)), GaN crystals are not regularly faceted and are strongly disoriented with respect to the c-sapphire substrate. By increasing the temperature to 950 °C (see Fig. 2(b)), the orientation along the c-axis is improved, but the shape is still irregular. From 1000 to 1040 °C (see Fig. 2(c)-(d)), we observe more regular hexagonal shapes composed of flat $\langle 0\,0\,0\,1 \rangle$ horizontal planes on top (*i.e.* prisms) and six $\langle 1\,\bar{1}\,0\,0 \rangle$ vertical side facets. Despite the significant improvement of the crystalline quality and orientation with increasing growth temperature, the homogeneity of the nucleation in the mask openings is strongly degraded since only a few openings are filled for 1040 °C growth. Actually, the percentage of opening fillings noticeably decreases from nearly 85 % at 900 - 950 °C, to 60 % at 1000 °C and to about 20 % at 1040 °C. The SAG of GaN nanostructures on patterned sapphire substrates is therefore highly sensitive to the growth temperature in terms of crystal shape and homogeneity of nucleation selectivity: the formation of regular hexagonal nanostructures is obtained at high temperature with poor selectivity efficiency, while the opposite trends are observed for lower temperatures.

*3.1.2. Two-step temperature growth*

To combine high homogeneity of nucleation selectivity and high crystal quality of GaN nanostructures grown on sapphire, a two-step temperature process has been developed. The initial nucleation step is first made at 950 °C for 200 s with TMGa and ammonia sources to get a high homogeneity of the GaN nucleation seeds. The TMGa source is then stopped and the temperature is ramped from 950 °C to 1040 °C in 200 s under ammonia flow. Finally, the TMGa supply is restarted for 350 s at higher temperature (1040 °C) to grow high quality GaN crystals. The other growth parameters (pressure, flows of carrier gas and precursors) remain the same as written in Sect. 2. Fig. 3(a)-(c) shows the 45°-tilted SEM measurements after each step of the growth. The evolution of GaN crystal shape is schematically illustrated in Fig. 3(d).

As observed in Fig. 3(a), GaN growths at 950 °C result in c-oriented crystals with mostly an irregular truncated pyramidal shape (see the inset) with a filling percentage of mask openings around 85 %. The average diameter of the GaN nanostructures is about 500 nm that means there is already a lateral overgrowth outside the 400 nm openings. The temperature ramping under ammonia changes the crystal shape from irregular truncated pyramids to well-shaped hexagons as shown in the inset of Fig. 3(b) while keeping the homogeneity of nucleation selectivity. This anisotropic shape evolution of GaN nanostructures may be explained first by a mass transport driven by Ga diffusion to reach the energetically stable facets, i.e. the ($1\bar{1}00$) planes at high temperature for growth on pre-nitridated sapphire substrates, which gives rise to the facets of the hexagons [20, 21]. In addition, the isotropic size reduction of the hexagonal nanostructures compared to the initial truncated pyramids in both lateral and vertical directions is also observed and could be explained by a decomposition of GaN at high temperature induced by $H_2$ etching coming from the $NH_3$ decomposition. The duration of the step is thus critical to avoid the complete vanishing of GaN seeds. Notice, the exact value of this duration should be optimized as a function of MOCVD setup. In the third step, the restart of the GaN growth maintains the previous hexagonal shape (see Fig. 3(c) and the inset). The typical diameter of hexagonal nanostructures after a supplementary growth duration of 350 s is about 500-700 nm in the lateral dimension with a 0.6-1 aspect ratio. The increase in size distribution of the GaN nanostructures with the growth time is probably due to the uncontrolled lateral overgrowth on the mask and to the initial size distribution of the nucleation seeds.

To sum up the multi-step temperature growth approach, the first step allows to get high homogeneity of nucleation selectivity of the seeds, the annealing step at ramped temperature induces the formation of hexagonal islands with uniform size (400 ± 20 nm) and the third step leads to hexagonal prismatic nanostructure growth.

*3.2. Discussion of growth mechanism*

*3.2.1. Shape evolution versus growth temperature*

In the last decade, the relationship between the crystal shape of GaN nanostructures and the growth parameters, especially the temperature has been mainly investigated by using the Wulff's plot (γ-plot) which deals with the anisotropy of the surface energy. It has been found in molecular beam epitaxy SAG that the *equilibrium* geometry of GaN nanostructures may change from pyramid to hexagon by decreasing the growth temperature due to the facet energy anisotropy [22]. In the case of MOVPE SAG process, Tanaka *et al*. [13] have reported the modification of SAG stripe morphologies from trapezoidal to rectangular cross-section with an increase of growth temperature from 850 to 1040 °C on pre-nitridated sapphire substrates. The *kinetic* influence in terms of kinetic Wulff's plot (*v*-plot) must also be taken into account in a real MOVPE growth condition [23, 24]. For instance, it has been found that the GaN crystal polarity strongly influences the growth rate of different crystal facets at high temperature leading to hexagonal N-polar wires [17]. In conclusion, the observed shape evolution of the GaN nanostructures from pyramidal shape at low temperature to hexagonal prismatic shape at high temperature can be understood by a combination of the influences of temperature-related facet energy anisotropy and kinetics. Further experiments to quantify this behaviour in details are under way.

*3.2.2. Selectivity versus growth temperature*

The MOVPE growth rate is in general determined by the competition between deposition from the vapour phase transport and desorption phenomena (evaporation, $H_2$ etching...). In our growth temperature range (above 950 °C), the transport of species in the vapour phase can be considered as constant (transport regime), whereas the desorption rate increases with the growth temperature (desorption regime). Additionally, the desorption of N atoms should be higher than in the standard two-dimensional GaN growths because of the low V/III ratio used to avoid the lateral growth, when a high ammonia partial pressure is generally required above 1040 °C to limit N-desorption [25]. In the case of SAG, the effective diffusion length of the reactive species on the mask (called the migration length) is a third effect that contributes significantly to the growth rate [26, 27, 28]. The migration length is usually strongly dependent on the growth temperature [29, 30]. At high temperature, ammonia molecules are immediately decomposed to adducts in the gas phase [25], and therefore, mainly Ga species are involved in the mask migration process. By increasing temperature, the migration length of the Ga species is decreased due to the significant reduction of the residence time of Ga species on the mask, which reduces the amount of supplied materials for incorporation [30]. Thus, the combination of deposition, desorption and migration contributions leads to a reduction of the nanostructure growth rate when the growth temperature is increased in the 900-1040 °C range. Furthermore, we can suspect that the nucleation of GaN crystals does not occur exactly at the same time in all openings due to the different surface states in terms of chemistry and structure: residues of resist, different diffusion barriers, surface defects, efficiency of nitridation (with respect to polarity control [17]). The seeds, which appear first, will serve as "sink" to collect the contiguous growth species. It then creates an adatom concentration gradient on the nearby surface that may lead to a diffusion network [31]. For 1040 °C growth, less material is provided compared to the growths at lower temperature. Therefore, the quantity of provided material is not large enough to compensate the desorption effect in the mask openings, except for some sites benefiting from the "sink" effect that may locally increase the growth rate. It finally gives a poor homogeneity of nucleation selectivity of SAG growth at high temperature. In the case of growths at lower temperature (900 and 950 °C), although the "sink" effect still exists, the growth rate is high enough to fill most of the mask openings, which leads to a higher homogeneity of nucleation selectivity.

For two-step temperature growth, the GaN seeds nucleate in most of mask holes due to low temperature process (950 °C) giving a high homogeneity of nucleation selectivity. After the annealing step under ammonia, the homogeneity of nucleation selectivity is maintained with a reduction of size distribution of nanostructure growth. Thanks to the high homogeneity of nucleation selectivity and size uniformity after the ammonia annealing, the "sink" effect corresponding to a local gradient of incoming species is strongly reduced at high temperature (1040 °C), and a SAG of GaN hexagonal prisms with high homogeneity of nucleation selectivity can be obtained.

## 4. Structural and optical characterizations of GaN nanostructures

*4.1. Crystalline quality of nanostructures*

Standard XRD measurements in the symmetric $\theta$-$2\theta$ geometry have been performed to investigate the crystallographic phases of the as-grown nanostructures obtained by the two-step growth process. As shown in Fig. 4, three Bragg peaks can be indexed with the (0 0 6) α-Al$_2$O$_3$ reflection ($R\bar{3}c$, c =1.2991 nm), the (0 0 2) and the (0 0 4) wurtzite GaN reflections ($P6_3mc$, c = 0.5188 nm). It indicates c-axis oriented growth of nanostructures on c-sapphire with no other secondary phases. To completely characterize the epitaxial relationships and the local structural properties, TEM measurements were performed on as-grown samples prepared by standard polishing and ion-milling processes. Fig. 5(a) and (b) depicts cross-section dark-field TEM images taken in a two-beam condition with g-vectors along <0 0 0 2> and <1 $\bar{1}$ 0 0> to image screw and edge dislocations having <0 0 0 2> and <1 1 $\bar{2}$ 0 > Burgers vectors. In the TEM images, a lateral overgrowth of the nanostructures outside the mask openings (see the limits indicated by the two black arrows in Fig. 5(a)) can be observed since the 550 nm nanostructure size is larger than the 400 nm mask opening. In the central part, we observe fringes in TEM images (Fig. 5(a)) coming from the large GaN/sapphire lattice mismatch and also misfit dislocations at the interface due to the strain relaxation. Threading dislocations (TD) are rarely observed: only two TDs with screw type are for instance visible in Fig. 5(a) (see white arrows) and no edge-type TD is observed in Fig. 5(b). On the contrary, in the lateral overgrowth part, a better crystal quality is observed with a low strain (absence of fringe) and no observed dislocations. Only a few stacking faults have been seen in Fig. 5(b) indicated by the white dashed triangular. It has been recently published that GaN nanorods grown on c-sapphire can present a high TD density [18]. These observations are attributed to the use of a thicker dioxide mask (100 nm) [18]. This evidences that a very thin mask (5 nm) significantly improve the crystalline quality of the SAG GaN nanostructure on sapphire. SAED pattern taken in the central interface area with the beam parallel to the $<1\ 1\ \bar{2}\ 0>$ direction of the GaN crystal is given in Fig. 5(c). The pattern indexation (black and white arrows denoting GaN and sapphire respectively) confirms the usual in-plane epitaxial relationship between GaN and sapphire ($[1\ \bar{1}\ 0\ 0]_{GaN}\ //\ [1\ 1\ \bar{2}\ 0]_{Sapp.}$) with a measured lattice mismatch of about 16.0 % along the $[1\ 0\ \bar{1}\ 0]$ direction, in agreement with the 16.09 % bulk value.

*4.2. Optical studies of single nanostructures*

Cathodoluminescence measurements in area-scanning mode were carried out at 5 K on single nanostructures to study the optical properties. Fig. 6 shows the CL spectra taken in a top view of single nanostructures (grown with the two-step temperature method) having 250, 500 and 1300 nm diameters and heights. The spectra are normalized to the intensity maximum. Fig. 7 gives the SEM image and the corresponding CL mappings measured at 3.46 and 2.26 eV in the top and cross-section views. In these images, darker area corresponds to larger emission intensity.

CL spectra exhibit two main contributions: the near band edge (NBE) emission of GaN in the UV region (~3.4 eV) and the yellow band (YB) emission at ~2.2 eV in the visible wavelength range, which is generally assigned to crystal defects like Ga vacancies and O complexes [32]. The relative intensity ratio between YB and NBE contributions strongly depends on the diameter of single nanostructures: it increases with the decrease of the nanostructure size as clearly observed in Fig. 6(a). To understand this behaviour, CL mappings were performed to localize the spatial emission of the NBE and YB wavelengths in the nanostructures. Fig. 7(a) shows the top view SEM image of the nanostructure arrays. The corresponding CL mapping images of the NBE and YB emissions at 3.46 and 2.26 eV are shown in Fig. 7(b) and (c) respectively. We can deduce from these measurements that the NBE emission mostly comes from the border part of the nanostructures, especially for the ones with size larger than 400 nm. On the contrary, the luminescence of the YB emission is more or less uniformly distributed whatever the nanostructure size. This observation is confirmed in Fig. 7(e) and (f) by CL mappings at 3.46 and 2.26 eV on 550 nm ion-milled sample in cross-section (the dashed line indicates the contour of the single nanostructure). This is consistent with the fact that the NBE emission is more sensitive to carrier diffusion than the YB emission, and the side part of the nanostructures is of much higher structural quality than the central part (see TEM results above). The size dependence of the optical properties can be also attributed to the fact that crystals with larger size have larger volume ratio between the side and the central part of the nanostructures leading to much stronger NBE emission. This result is in favour of the use of small diameter openings for optoelectronic applications.

In addition, we observe a variation of the NBE peak position of the nanostructures with different diameters as shown in Fig. 6(b). The peak shift can be assigned to two opposite effects: a blueshift due to compressive strain induced by the lattice mismatch [33] and a redshift induced by silicon doping [34]. Such unintentional Si doping, which is mostly located at the bottom part of the nanostructures, may come from the $Si_3N_4$ dielectric mask through a surface diffusion process as already reported in SAG pyramids [35]. Depending on the concentration, Si doping may induce a gradual redshift up to ~40 meV due to band gap renormalization [34, 35]. For nanostructures with diameter equal to 1300 nm, the NBE emission is principally related to the strain-relaxed lateral overgrowth. Thus, the NBE peak position located at 3.46 eV corresponds to the value of the fully relaxed GaN 2D layers at 5 K [36]. For nanostructures with a diameter close to 500 nm, the NBE emission is mostly attributed to the strain relaxed lateral overgrowth on the mask and the NBE redshift of 35 meV can be mainly assigned to the Si incorporation from the mask surface diffusion. For nanostructures equal to 250 nm, the NBE emission comes from the crystals grown in direct epitaxy with the sapphire since the nanostructures size is smaller than the mask openings. Thus, the compressive stain must be also taken into account in addition to the Si incorporation effect. The 28 meV blueshift of the NBE peak position compared to that of the nanostructure with 500 nm size can be assigned to this compressive strain. Moreover, the peak profiles of the NBE emission for nanostructures with a diameter of 500 and 250 nm are consistent with the Si doping assumption: The shape of the NBE emission is typical of Si-doped GaN, featuring a sharp cut-off at high energy and a low energy tail [34, 37]. From the observed FWHM of 60 meV, the doping level can be estimated to $10^{19}/cm^3$ [38].

## 5. Conclusions

In summary, a wafer-scale and effective SAG of GaN nanostructures on patterned c-sapphire substrate using NIL has been demonstrated. The temperature influence on the nanostructure shape has been studied for a single-step temperature growth: for temperature smaller than 1000 °C, a pyramidal shape is observed whereas a prismatic one is favored at higher temperature. In parallel, the SAG homogeneity of nucleation selectivity is significantly degraded at high temperature (> 1000 °C) due to a combination of several temperature-dependent effects, such as the desorption, the precursors migration length on the mask and the 'sink' effect due to the non-uniform GaN seed nucleation. Based on these results, SAG of hexagonal prismatic GaN nanostructures with high homogeneity of nucleation selectivity has been achieved using a two-step temperature process consisting of a low temperature nucleation at 950 °C followed by an ammonia annealing and a further growth at high temperature (1040 °C). Structural characterizations of GaN nanostructures show (i) a central zone in direct epitaxy with c-sapphire with a relatively poor crystal quality exhibiting misfit and threading dislocations and (ii) an overgrown area on the $Si_3N_4$ mask showing good crystalline quality. The CL mappings depict a relatively larger intensity of NBE emission from the border part of the nanostructures (*i.e.* overgrown on $Si_3N_4$) in agreement with a better crystalline. Consequently, it results in a decrease of the YB/NBE intensity ratio with an increase of the nanostructure size. Finally, the NBE peak shift of GaN prisms as a function of size is attributed both to the compressive strain and to the silicon incorporation through surface diffusion from the $Si_3N_4$ mask.

**Acknowledgments**

The authors thank J. Dussaud and M. Lafossas for technical support, F. Levy and P. Gilet from LETI/DOPT for patterned sapphire substrates. G.P.M acknowledges P.-H. Jouneau and G. Feuillet for scientific support and X.J.C. thanks for the financial support of the foundation « Nanosciences aux limites de la Nanoélectronique ». The work was partly funded by the French ANR Bonafo (ANR-08-Nano-031-01) and Sincrone (ANR-09-MAPR-0011-03).

**Figures**

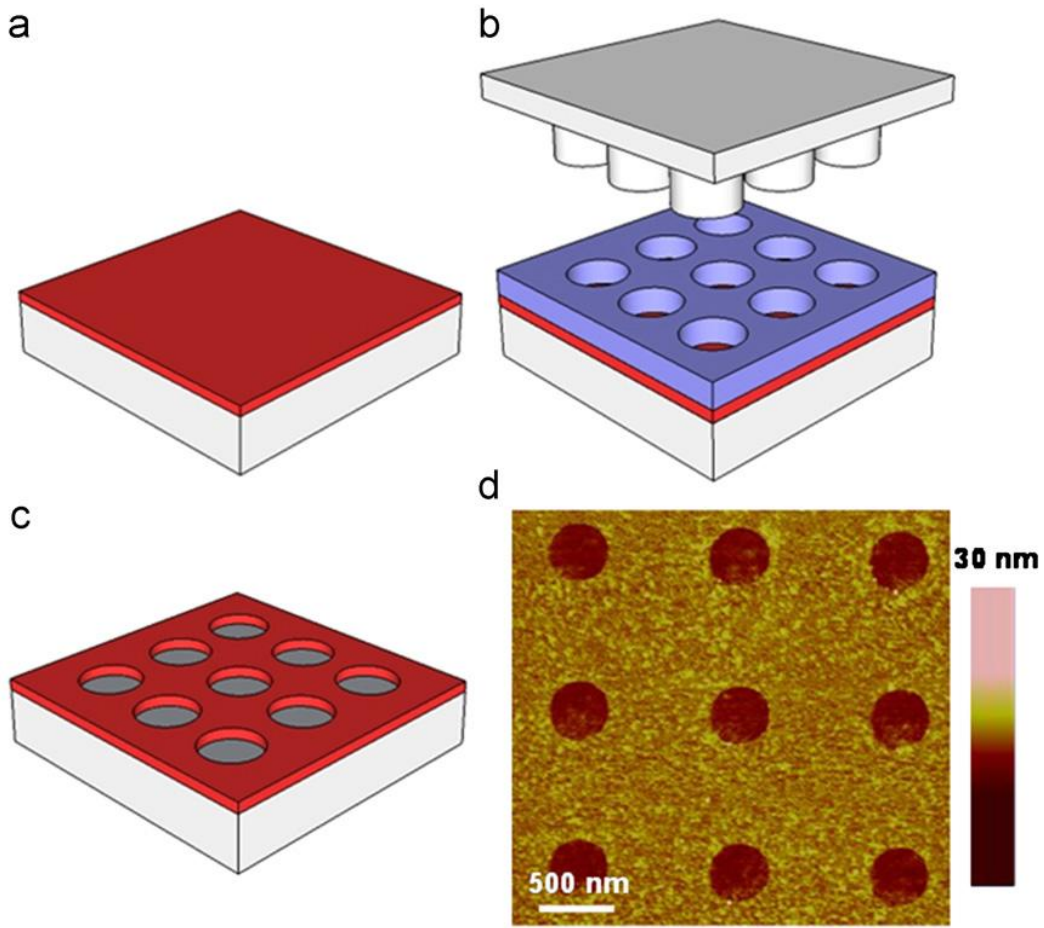

Fig. 1. (a)-(c) Schematic of the nanoimprint patterning process on c-sapphire substrate. (d) Typical AFM image of the patterned $Si_3N_4$ mask: array of 400 nm holes spaced by 1.1μm.

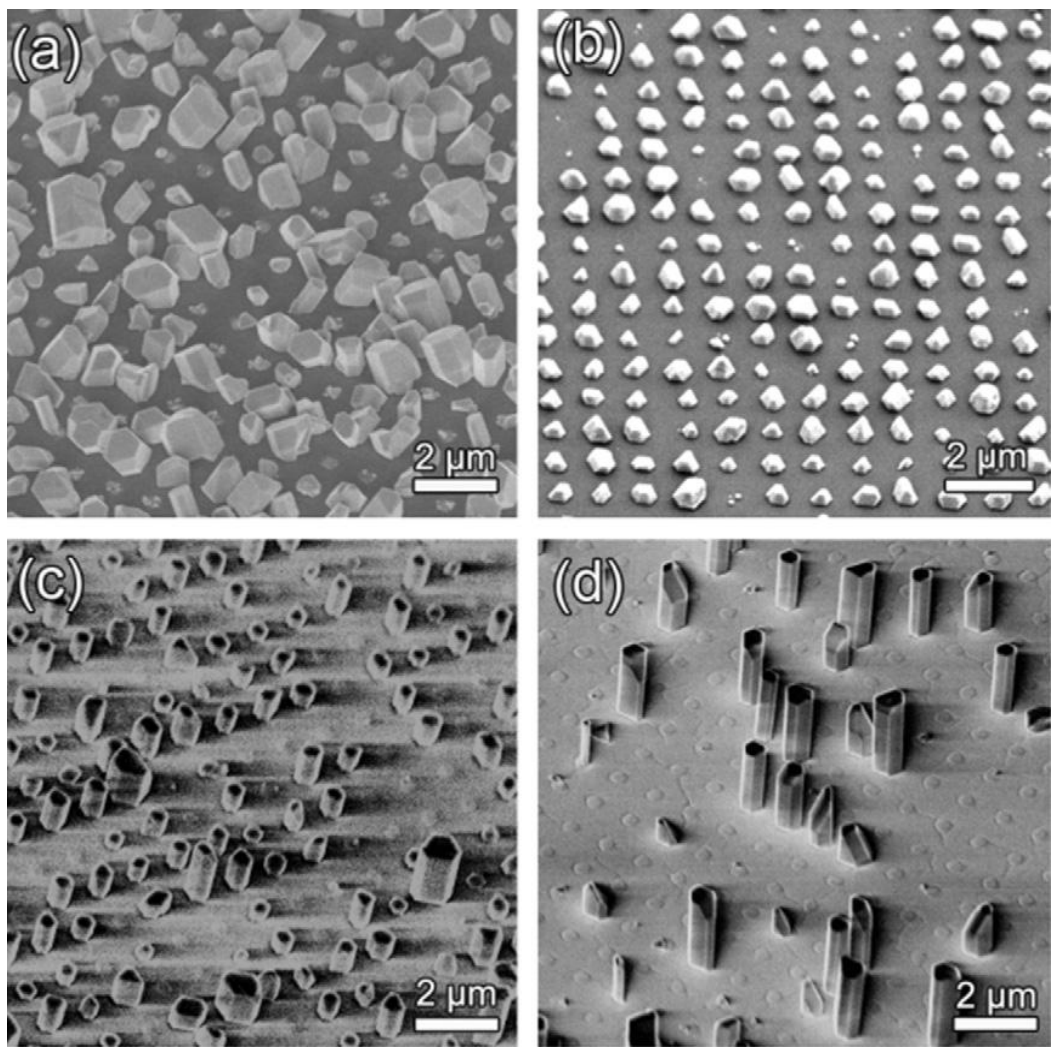

Fig. 2. 45º-tilted SEM images of the self-assembled growth (SAG) for a single-step temperature process at (a) 900 ºC, (b) 950 ºC, (c) 1000 ºC and (d) 1040 ºC.

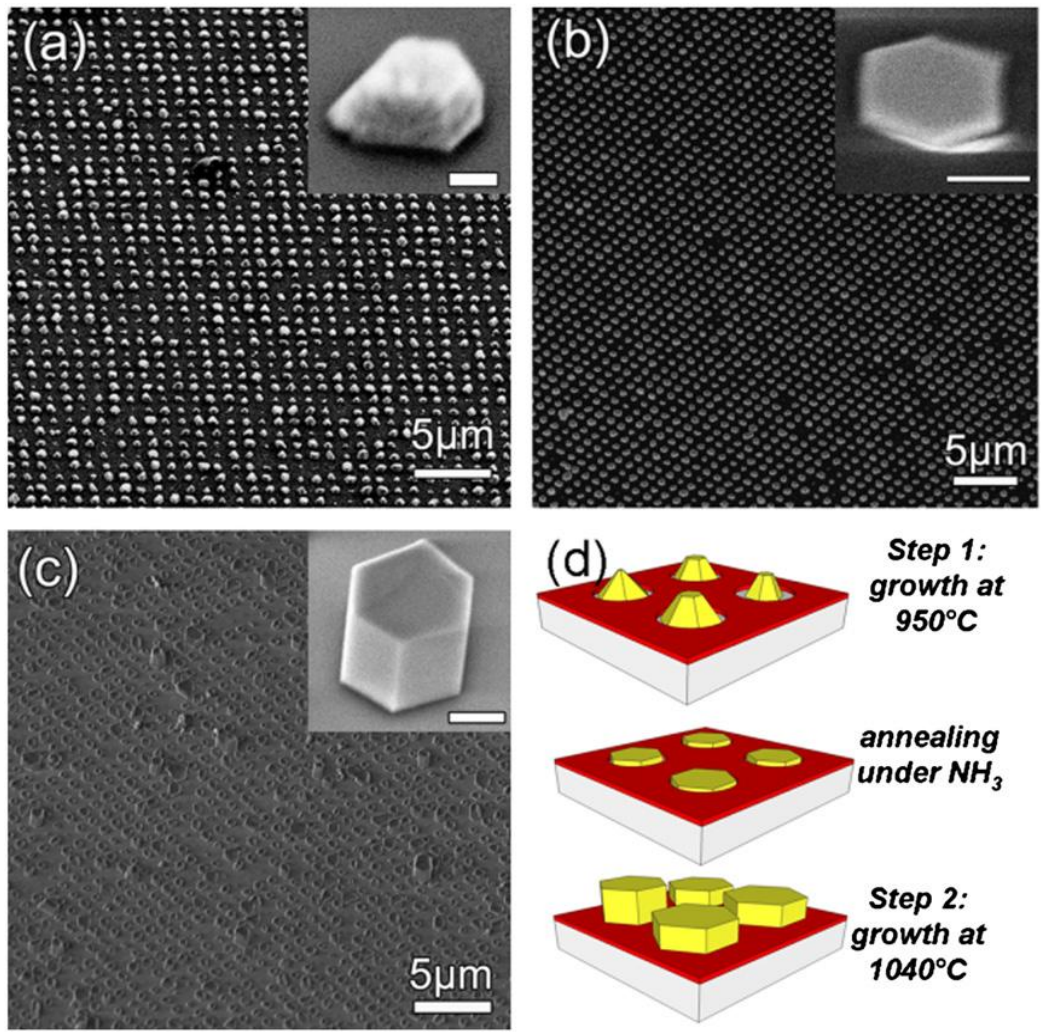

Fig. 3. 45°-tilted SEM images of the major stages of the two-step temperature process for the GaN SAG on c-sapphire: (a) after 200 s GaN deposition at 950 ºC, (b) after 200 s temperature ramping from 950 ºC to 1040 ºC under ammonia, (c) after 350 s GaN deposition at 1040 ºC. The insets of each image give an enlarged view of typical single objects in the openings (the inset scale bars are 200 nm). (d) Schematic describing the shape evolution of GaN nanostructures during this two-step temperature growth method.

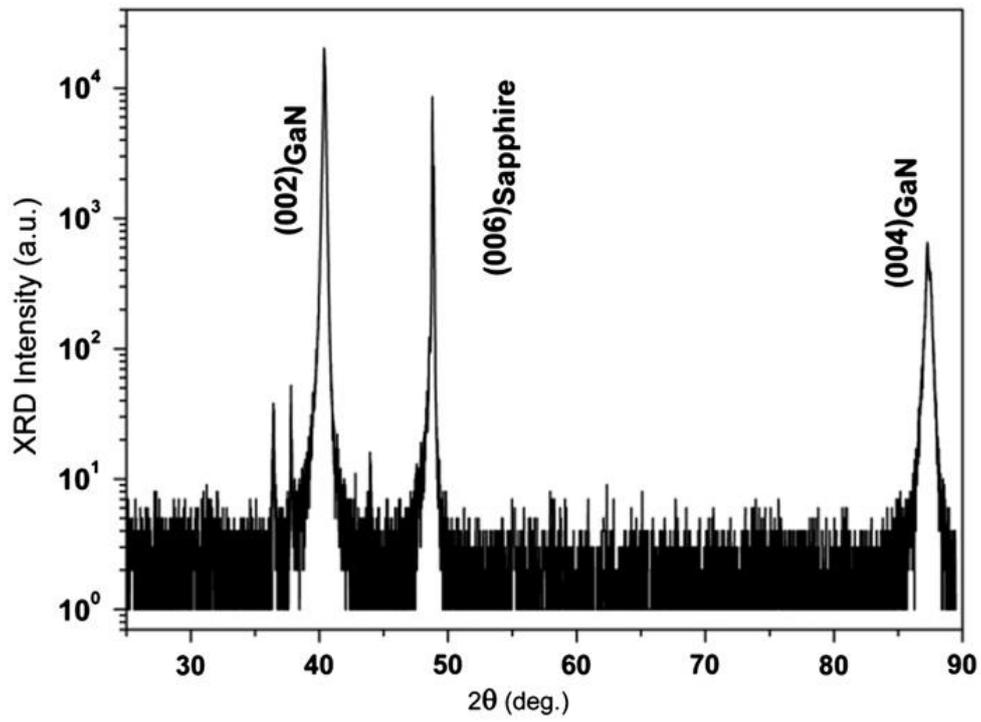

Fig. 4. X-ray diffraction spectrum (standard θ-2θ geometry, λ=0.17902 nm) of as-grown GaN nanostructure arrays after the two-step temperature growth method described in the text.

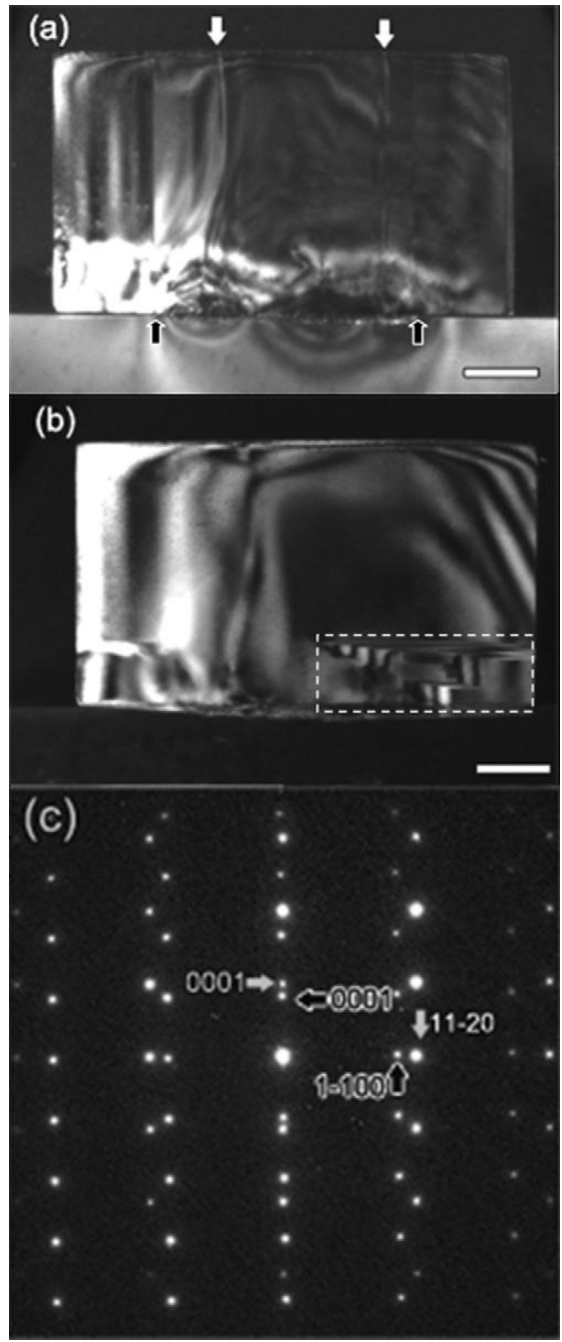

Fig. 5. (a) and (b) Cross-section dark field TEM images of 500 nm GaN hexagonal nanostructures with <0 0 0 2> and <10 $\bar{1}$ 0> g-vectors, respectively. White arrows in (a) denote threading dislocations (TDs) and black arrows the mask limits. White dashed rectangular in (b) denotes the stacking faults (SFs). The scale bars are 100 nm. (c) Electron diffraction pattern taken at the GaN/Sapphire interface with a beam direction parallel to [1 1 $\bar{2}$ 0]$_{GaN}$. The spots indexed with white and black arrows correspond to sapphire and GaN, respectively.

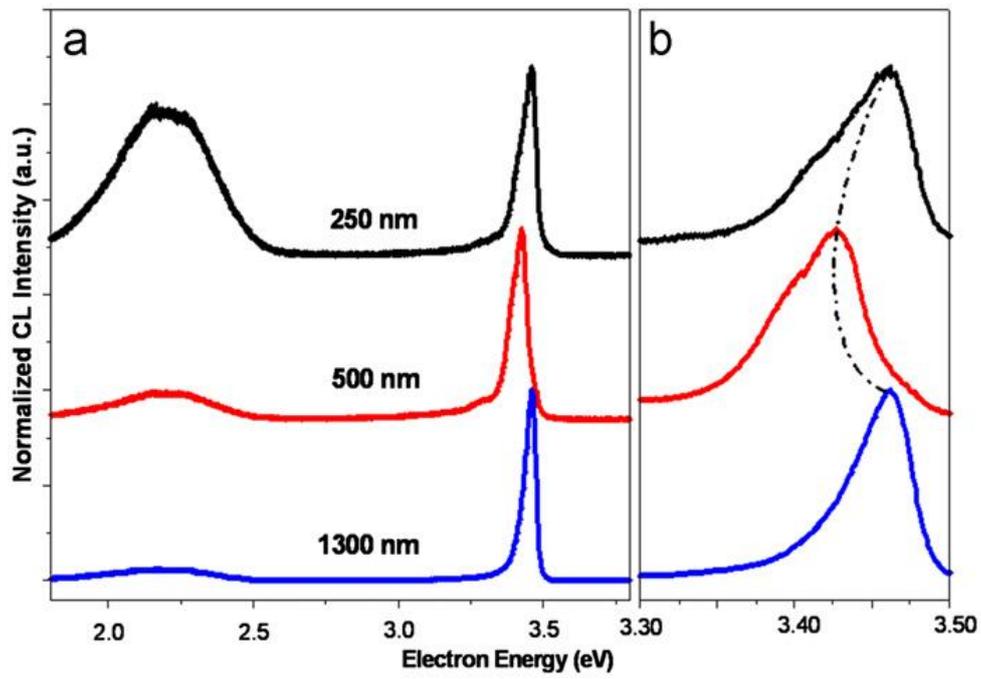

Fig. 6. (a) Normalized cathodoluminescence spectra at 5 K for single GaN nanostructures with 250, 500 and 1300 nm diameters. (b) Enlarged view corresponding to the near band edge emission contribution.

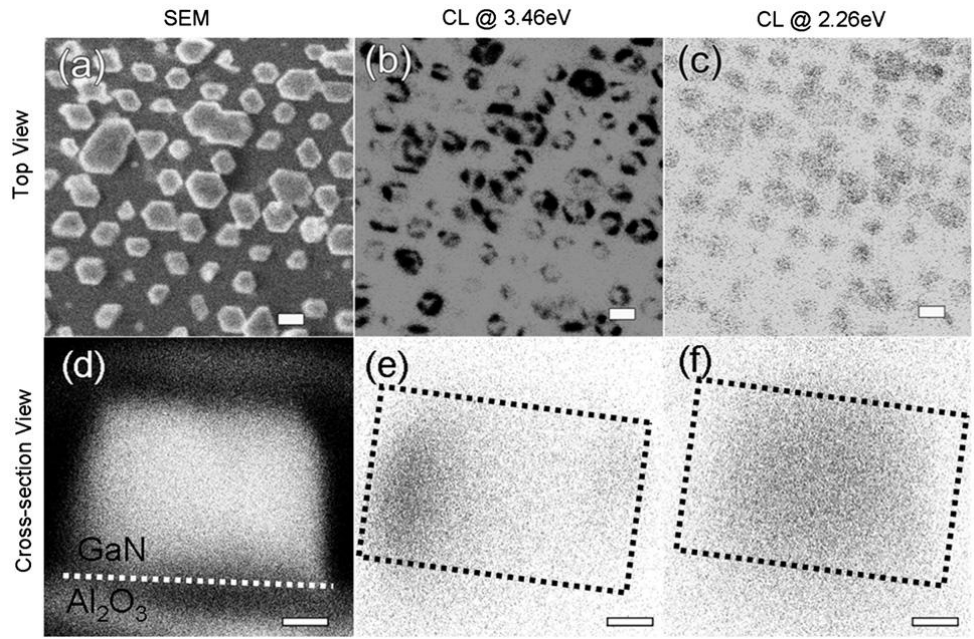

Fig. 7. (a) Top view SEM image of an array of GaN nanostructures grown by the two-step temperature method described in the text. Cathodoluminescence (CL) mappings at 5 K of the area shown in (a) at (b) 3.46 and (c) 2.26 eV respectively corresponding to the near band edge (NBE) and yellow band (YB) emissions. (d) Cross-section SEM image of a single GaN hexagonal prismatic nanostructure prepared by ion-milling. (e) and (f) correspond to CL mappings at 5 K of the nanostructure shown in (d) at 3.46 and 2.26 eV respectively. The dashed rectangles in (e) and (f) show the contour of the single object. Scale bars are 500 nm in (a)-(c) and 100 nm in (d)-(f).